\documentstyle[11pt,newpasp,twoside,epsf]{article}
\def\msun{M$_{\odot}$}
\def\etal{{\it et al.} }
\newcommand{\psr}{PSR~J1740-5340~}

\def\edcomment#1{\iffalse\marginpar{\raggedright\sl#1\/}\else\relax\fi}
\reversemarginpar

\begin{document}
\title{To accrete or not to accrete: the dilemma of the recycling scenario}
 \author{L. Burderi, F. D'Antona}
\affil{Osservatorio Astronomico di Roma, via Frascati 33, 00040 Roma, Italy}
\author{T. Di Salvo, G. Lavagetto, R. Iaria, N.R. Robba}
\affil{Dipartimento di Scienze Fisiche ed Astronomiche, Universit\`a di
Palermo, via Archirafi 36, 90123 Palermo, Italy}

\begin{abstract}
We study the evolution of a low-mass X-ray binary by coupling a binary 
stellar evolution code with a general relativistic code that describes the 
behaviour of the neutron star. We find that non-conservative mass transfer
scenarios are required to prevent the formation of submillisecond pulsars
and/or the collapse to a black hole. We discuss the sweeping effects of an
active magneto-dipole rotator on the transferred matter
as a promising mechanism to obtain highly non-conservative evolutions.
\end{abstract}

\section{Introduction: a review of the Recycling Scenario}

The ``classical'' scenario for the formation of millisecond radio
pulsar binaries
with low-mass companions envisages four main stages (see {\it e.g.}
Bhattacharya \& van den Heuvel 1991 for a review): (i) The magnetic moment
$\mu$ of the newly formed radio pulsar ($\mu = B R^{3}$, where $B$ and
$R$ are the surface magnetic field and the neutron star, NS, 
radius\footnote{Actually, since the NS is a relativistic object, $R$ is 
the NS circumferential equatorial radius, i.e.\ the proper circumference 
in the equatorial plane divided by $2 \pi$.}, respectively) decays 
spontaneously on a e-folding timescale $t_{\mu} \sim 10^{7} - 10^{8}$ yr 
(e.g.\ Lyne, Anderson \& Salter 1982) from an initial value
$\sim 10^{31}$ G cm$^3$ to a final value $\sim 10^{26} - 10^{27}$ G cm$^3$,
when the decay probably stops
(e.g.\ Bhattacharya \& Srinivasan 1986).
(ii) Simultaneously, the pulsar spin period $P$ increases
under magnetic dipole emission according to $L_{\rm PSR} = (2/3c^{3}) \mu^{2}
(2 \pi/P)^{4}$ (where $L_{\rm PSR}$ is the bolometric magneto--dipole
luminosity and $c$ is the speed of light).
The pulsar swithces off when eventually crosses the ``death
line'' in the $\mu-P$ plane, {\it i.e.} the line defined by the relation
$\mu_{26}/P^2 = 2 \times 10^{3}$ (where 
$\mu_{26} = \mu / 10^{26}$ G cm$^3$) below which it is believed that
the radio pulsar phenomenon does not take place (see {\it e.g.} Ruderman \& 
Sutherland 1975).
(iii) The companion star overflows its Roche lobe and
transfers mass with angular momentum to the NS {\it via} a Keplerian accretion
disc, thereby spinning it up to millisecond periods (close to the Keplerian 
period at inner rim of the accretion disk, see below) and back across the 
death line (recycling). During this phase the system is visible as a 
low-mass X-ray binary (LMXB). 
(iv) Mass transfer ceases: the NS is again visible as a radio pulsar whose
spin rate decays under magnetic dipole emission very slowly as $L_{\rm PSR}
\propto \mu^{2}$ and $\mu$ is reduced by three or four orders of magnitude.
The end point is therefore a millisecond pulsar orbiting a low mass
companion ($< 0.3$ \msun)
that is the remnant of the $\sim 1$ \msun\ mass donor.

\subsection*{Recycling in Transient Systems}
Most LMXBs are NS Soft X-ray Transients (NSXT), i.e.\ transient 
systems harboring a NS (see Campana \etal 1998 for a review).  
Adopting the same conversion efficiency of the accreting
matter energy into X-rays during the outbursts and the quiescent
states (but see Barret \etal 2000 for a different explanation), the 
inferred variations in the accretion rate are a factor $\sim 10^{5}$.

NSXTs can provide a direct evidence of the recycling scenario, since
during stage (iii) the mass transfer rate varies up to five orders of
magnitude. During the LMXB phase the accretion disk is truncated because 
of one of the following reasons: 
(i) the interaction with the magnetic field of the NS, which 
truncates the disc at the magnetospheric radius $R_{\rm M}$,
at which the accretion flow is channeled along the magnetic field 
lines towards the magnetic poles onto the NS surface; 
(ii) the presence of the NS surface itself at $R$; and (iii) the lack of 
closed Keplerian orbits for radii smaller than the marginally stable orbit 
radius, $R_{\rm MSO}$ (at few -- depending on the mass and spin of the compact 
object -- Schwarzschild radii from the NS centre). The position
of $R_{\rm M}$ is determined by the istantaneous balance of the
pressure exerted by the accretion disc and the pressure
exerted by the NS magnetic field: 
\begin{equation}
R_{\rm M} = 1.0 \times 10^{6} \, \phi\ \mu_{26}^{4/7} m^{-1/7} R_6^{-2/7}
\dot{m}^{-2/7}
\; \; \; {\rm cm}\ ,
\label{eq:rma}
\end{equation}
where $\phi \le 1$, 
$m$ is the NS gravitational mass in \msun\footnote{Actually, since 
the NS is a relativistic object, it is important to distinguish between 
the baryonic mass, roughly speaking a measure of the amount of matter, 
and the gravitational mass that is smaller by a factor of 
$\sim (3/5) G M^2/R \sim 0.1 M c^2$, corresponding to the 
binding energy of the NS.}, $R_{6}$ is the NS radius in units of 
$10^{6}$ cm, and $\dot{m}$ is the baryonic mass accretion
rate$^2$ in Eddington units (the Eddington accretion rate
is $1.5 \times 10^{-8} R_{6} \; \; \; {\rm M}_{\odot} {\rm yr}^{-1}$).
Eq. 1 
shows that as $\dot{m}$ decreases, $R_{\rm M}$ expands.

Accretion onto a spinning magnetized NS is centrifugally inhibited once
$R_{\rm M}$ expands beyond
the corotation radius $R_{\rm CO}$, at which the Keplerian
angular frequency of the orbiting matter is equal to the NS
spin:
$R_{\rm CO} = 1.5 \times 10^{6} \, m^{1/3} P_{-3}^{2/3} \; \; \; {\rm cm}$
where $P_{-3}$ is the NS spin period in milliseconds.
In this case the accreting matter could in
principle be ejected from the system: this is called propeller phase
(Illarionov \& Sunyaev 1975).
Finally, if $R_{\rm M}$ further expands beyond the  light-cylinder
radius (where an object corotating with the NS attains the speed
of light,
$R_{\rm LC} = 4.8 \times 10^{6} \, P_{-3}  \; \; \; {\rm cm}$),
the NS becomes generator of magnetodipole
radiation and relativistic particles.
Indeed,  a  common requirement  of  all  the  models of  the  emission
mechanism  from   a  rotating  magnetic  dipole  is   that  the  space
surrounding the NS  is free of matter up to $R_{\rm LC}$.

Let us consider the behaviour of a NSXT at the end of an 
outburst. Adopting $\dot m \sim 1$ in outburst,
eq. 1 
gives
$R_{\rm M} \sim R \sim 10^{6} \; \; {\rm cm}$.
In quiescence $\dot{m} \sim 10^{-5}$ and 
$R_{\rm M}= (10^{-5})^{-2/7} \times 10^6 = 2.7 \times 10^{7} \; \; {\rm cm}\ 
> R_{\rm LC}$,
for spin periods up to few milliseconds. Therefore it is likely that, during 
the quiescent phase, a magneto-dipole emitter switches on (see {\it e.g.} 
Stella \etal 1994; Burderi \etal 2001).
In this case, as the NS $\mu$ and $P$ place such a system above the
``death line'', it is plausible to expect that the
NS turns-on as a millisecond radio pulsar until a new outburst episode
pushes $R_{\rm M}$ back, close to the NS surface, quenching radio
emission and restoring accretion.

\section{To accrete: conservative mass transfer}
A rapidly rotating NS can support a maximum 
mass (against gravitational collapse) much higher than the non-rotating mass 
limit, since the centrifugal force attenuates the effects of the 
gravitational pull. Conversely, if a rotating NS has a mass that exceeds the 
non-rotating limit (supramassive NS), it will be subject to gravitational 
collapse if it loses enough angular momentum, $J$. 
In contrast to the standard behaviour, supramassive NSs spin-up just before 
collapse, even if they lose energy (Cook, Shapiro \& Teukolsky 1992).
The value of the maximum rotating and non-rotating mass depends on the 
equation of state (EOS) governing the NS matter, and the minimum allowed 
period for a given mass and EOS occurs when gravity is balanced by 
centrifugal forces at the NS equator (mass shedding limit). In this 
case the NS has the maximum $J$ allowed for that given mass,
$J_{\rm max}$.
The NS radius, which is always in the order of $10^6$ cm, depends on 
the mass, EOS, and $J$ of the NS: a rapidly 
rotating NS can have a much larger radius than a non-rotating one 
(up to 40\%, Cook, Shapiro \& Teukolsky 1994).

In order to study the evolution of a LMXB in detail we coupled 
a modified version of rns (rotating NSs) public domain code by Stergioulas 
\& Morsink (1999), that allowed us to build a complete grid of relativistic 
equilibrium configurations, with the binary evolution evolution
code ATON  (D'Antona, Mazzitelli \& Ritter, 1989). We assume the NS
to be low-magnetized ($B \sim 10^8$ G). Our computations show that during 
the binary evolution, the companion transfers as much as 1 \msun to the 
NS, with an accretion rate of $\sim 10^{-9}$ \msun yr$^{-1}$. 
This rate is sufficient to keep the inner radius of the accretion disc, 
$r_D$, in contact with the NS surface, thus preventing the onset of a 
propeller phase capable of ejecting a significant fraction of the matter 
transferred by the companion. 

Matter leaves the inner radius of the disk $r_D$ with a specific angular 
momentum $j$ that is a function of $r_D$, $M_G$, and $J$ (Bardeen 1970)
We can therefore write the evolutionary equations:
\begin{equation}
  \label{eq:j}
\left\{\begin{array}{lll}
d M_B / d t & = & \dot{M}_{C} \\
d J / dt & = & f(J) \cdot \dot{M}_B
\end{array}
\right.
\end{equation}
where $M_{B}$ is the baryonic mass of the star and $\dot{M}_{C}$ is
the mass transfer rate from the companion, and 
$f(J) = j$ for $J < J_{\rm max} (M_B)$ and $f(J) = d J_{\rm max} / d M_B$
for $J = J_{\rm max} (M_B)$, since we have assumed that at mass shedding
the matter of the disc dissipates angular momentum in order to accrete onto 
the NS, keeping the NS at the mass shedding limit.

Once the accretion ends, $R_{\rm M}$ expands beyond $R_{\rm LC}$ and 
a magneto-dipole rotator (radio pulsar) switches-on. 
In this situation is convenient to describe the NS evolution in 
terms of baryonic mass and total energy rather than in terms of baryonic 
mass and angular momentum: 
\begin{equation}
  \label{eq:evpul}
\left\{\begin{array}{lll}
   d M_{B} / dt & = & 0\\
   d M_{G} / dt & = & - (2/3c^{5}) \mu^{2} (2 \pi/P)^{4}
\end{array}
\right.
\end{equation}
We integrate the differential equations 2 or 3 using a 
finite-differences method (see Lavagetto \etal 2004 for details). 

\vskip 0.1cm
\begin{figure}
  \plottwo{aspen_f1.eps}{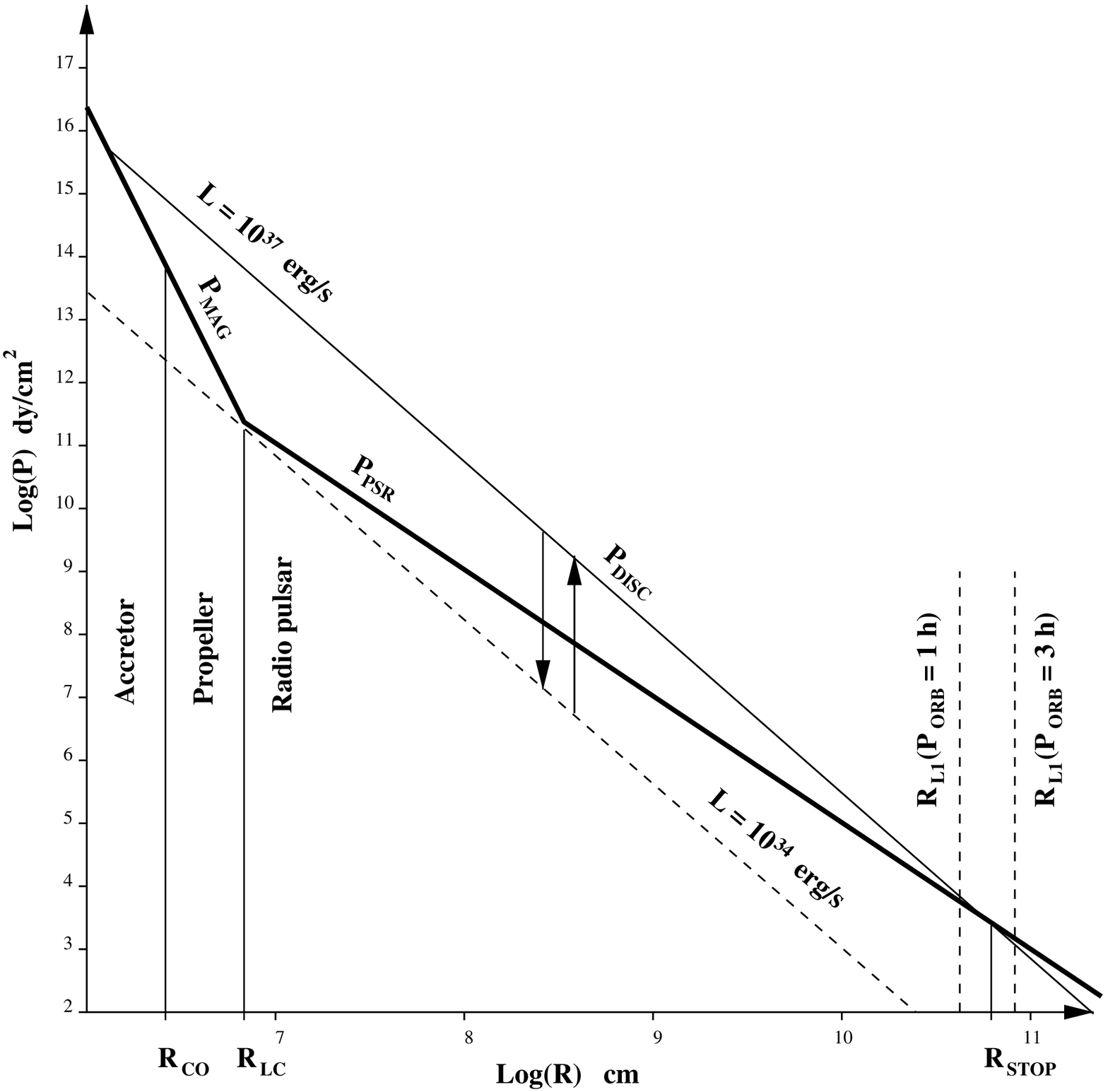}
  \caption{Left: Evolution of the NS in the mass-radius plane. Labelled dots 
  indicate the corresponding value of the NS spin period in ms.
  Right: Radial dependence of the pressures relevant for the evolution of 
  accreting NSs and recycled pulsars. The parameters adopted are $\mu_{26} 
  = 5$, $P_{-3} = 1.5$, $R_6 = 1$, $m = 1.4$}
\end{figure}

In Figure 1 (left panel) we plot, in the mass-radius plane, the 
evolution of a system with a donor star of $1.15 M_{\odot}$, 
which loses $0.91 M_{\odot}$ during the accretion phase, which is 
representative of most of the simulations. 
To describe the NS we adopted the quite standard EOS 
FPS. The upper dashed curve is the mass-shedding limit,
the lower dashed curve is the the sequence of stable non-rotating 
configurations, the thin solid line on the lower right is the limit beyond
which the NS collapses onto a black hole, the thick solid line shows
the evolution of the NS during the accretion phase, and, finally, the thick 
dashed line shows the subsequent radio-pulsar phase.
The first phase of accretion is charachterized by 
a rapid spin up that brings the star to the mass-shedding limit, along
which it evolves with $P$ slowly decreasing from 0.78 to  0.58 ms. 
Once the accretion stops during the radio-pulsar
phase, $P$ decreases (because the NS is supramassive) to 0.54 ms, 
until the NS eventually collapses to a black hole.

All of our simultations indicate that: a) for most EOS an accretion 
induced collapse to a black hole is almost unavoidable; 
b) in the absence of an external mechanism that brakes down the 
NS (such as gravitational waves), a small fraction of the accreted matter 
($\sim 0.1 M_\odot$) is sufficient to spin-up the NS to periods below one 
millisecond. In contrast, no submillisecond pulsars have been detected up 
to date: the shortest observed spin period is 1.5 ms (Backer \etal 1982), 
uncomfortably higher than the theoretical predictions.
Both these predictions are in contrast with the existence of the population
of MSP all spinning at frequencies well above 1 ms.

\section{Not to Accrete: non conservative mass transfer}

A non-conservative mass transfer has been often invoked to overcome the
difficulties outlined above. Indeed the ejection of more than 90\%
of the transferred mass could prevent the gravitational collapse and keep
the NS spin period above 1 ms. The propeller effect discussed in \S 1
is, in principle, a promising mechanism to obtain highly non-conservative
evolutions. However, the virial theorem sets stringent limits on the fraction 
of matter that can be ejected in this phase. In fact, it states that, at any 
radius in the disk, the virialized matter has already liberated (via 
electromagnetic radiation) half of its available energy. Considering that 
$r_D \sim R$ for these kind of systems, the accreting matter has already
radiated 50\% of the whole specific energy obtainable from accretion 
($\epsilon_{\rm acc} \sim G M_G/R$) before setting onto the NS surface. 
Therefore to eject this matter 
$\sim 0.5 \epsilon_{\rm acc}$ must be given back to it. As the only source 
of energy is that stored in the NS as rotational energy by the accretion 
process itself, the typical ejection efficiency is $\leq 50\%$, considering 
that once the system has reached the spin equilibrium, no further spin-up 
takes place and the storage of accretion energy in rotational energy is 
impossible. Thus, the accreted mass is $\ge 0.5$ \msun, well enough to
spin-up the NS to submillisecond periods.
An alternative viable hypothesis to explain the lack of ultrafast
rotating NSs is that gravitational wave emission balances the torque
due to accretion (see a review in Ushomirsky, Bildsten, \& Cutler
2000), although this effect cannot prevent the gravitational
collapse.

The only way to overcome these difficulties is to obtain ejection
efficiencies close to unity. This is indeed possible if the matter is
ejected so far away from the NS surface ($r >> R$) that it has an almost
negligible binding energy $\epsilon = G M_G /r << \epsilon_{\rm acc}$. 
As the NS is spinning very fast, the switch-on of a radio pulsar is 
unavoidable once $R_M > R_{LC}$. In this case the
pressure exerted by the radiation field of the radio pulsar may
overcome the pressure of the accretion disk, thus determining the
ejection of matter from the system. Once the disk has been swept away,
the radiation pressure stops the infalling matter as it overflows the
inner Lagrangian point $L_1$, where $\epsilon \sim 0$.

The push on the accretion flow exerted by the magnetic
field of the NS can be described in terms of an outward pressure
(we use the expressions {\it outward} or {\it inward pressures}
to indicate the direction of the force with respect to the radial direction):
$P_{\rm MAG} = B^2/4\pi = 7.96 \times 10^{14} \mu_{26}^2 r_6^{-6} \;\;
{\rm dy}/{\rm cm}^2
\label{eq:pmag} $,
where $r_6$ is the distance from the NS center in units of $10^6$ cm.
If the disk terminates outside $R_{\rm LC}$, the outward
pressure is the radiation pressure of the rotating
magnetic dipole, which, assuming isotropic emission, is
$P_{\rm PSR} = 2.04 \times 10^{12} P_{-3}^{-4}
\mu_{26}^2 r_6^{-2} \;\; {\rm dy}/{\rm cm}^2.
\label{eq:ppsr} $
In Figure~1 the two outward pressures ($P_{\rm MAG}$ and $P_{\rm PSR}$)
are shown as bold lines for typical values of the parameters
(see figure caption).

The flow, in turn, exerts an inward pressure on the field.
For a Shakura--Sunyaev accretion disc
(see {\it e.g.} Frank, King \& Raine 1992):
$P_{\rm DISK} = 1.02 \times 10^{16} \alpha^{-9/10}
n_{0.615}^{-1} L_{37}^{17/20} m^{1/40} R_6^{17/20} f^{17/5} r_6^{-21/8}
\;\; {\rm dy}/{\rm cm}^2
\label{eq:pgas}  $,
where $\alpha$ is the Shakura--Sunyaev viscosity parameter,
$n_{0.615} = n/0.615 \sim 1$ for a gas with solar abundances
(where $n$ is the mean particle mass in units of the proton mass),
and $f = [1 - (R_6/r_6)^{1/2}]^{1/4} \le 1$.
We measure $\dot M$ in units of $L_{37} = L/10^{37}$ ergs/s from 
$L = G M \dot M /R$.

In Figure~1 (right panel) the inward disc pressure
for a luminosity $L_{\rm MAX}$, corresponding to the outburst luminosity
of an NSXT, is shown as a thin solid line.
The disc pressure line, which intersects $P_{\rm MAG}$ at $R_{\rm LC}$,
defines a critical luminosity $L_{\rm switch}$, shown as a dashed line in 
Fig.~1, at which the radio pulsar switches-on. 

The intersections of the $P_{\rm DISC}$ line corresponding to $L_{\rm MAX}$
with each of the
outward pressure lines define equilibrium points between the inward and outward
pressures. The equilibrium is {\it stable} at $r = R_{\rm m}$, and {\it
unstable} at $r = R_{\rm STOP}$, which can be derived equating $P_{\rm PSR}$
and $P_{\rm DISK}$:
$R_{\rm STOP} \sim 8 \times 10^{11} \alpha^{-36/25}
n_{0.615}^{-8/5} R_6^{34/25} f^{136/25} 
L_{37}^{34/25} m^{1/25} \mu_{26}^{-16/5} P_{-3}^{32/5}  \;\; {\rm cm}$.
In fact, as $P_{\rm MAG}$ is steeper than $P_{\rm DISC}$, if a
small fluctuation forces the inner rim of the disc inward (outward),
in a region where the magnetic pressure is greater (smaller) than the
disc pressure, this results in a net force that pushes the disc back
to its original location $R_{\rm m}$. As $P_{\rm PSR}$ is flatter than
$P_{\rm DISC}$, with the same argument is easy to see that no stable
equilibrium is possible at $R_{\rm STOP}$ and the disc is swept away
by the radiation pressure.
This means that, for $r>R_{\rm STOP}$, no disc can exist for
any luminosity $\leq L_{\rm MAX}$.
The sudden drop in the mass-transfer rate during the quiescent phase of a
NSXT initiates a phase that we termed ``radio
ejection'', in which the mechanism that drives mass overflow
through $L_1$ is still active, while the pulsar radiation pressure prevents 
mass accretion.
As the matter released from the companion cannot accrete, it is now
ejected as soon as it enters the Roche lobe of the primary.
The distance of $L_1$ from the NS center, $R_{\rm L 1}$, depends only
on the orbital parameters. In the approximation given by Paczynski (1971), 
we can impose $R_{\rm STOP}/ R_{\rm L 1} = 1$ and solve for the orbital
period: 
\begin{equation}
P_{\rm crit} =  1.05 \times (\alpha^{-36}
n_{0.615}^{-40} R_6^{34})^{3/50} L_{36}^{51/25} m^{1/10} \mu_{26}^{-24/5}
P_{-3}^{48/5}  g (m, m_2)\;\;  {\rm h}
\label{eq:pcrit}
\end{equation}
where $g (m, m_2) = [1 - 0.462 ( m_2 / (m+m_2))^{1/3} ]^{-3/2} (m+m_2)^{-1/2}$,
and $m_2$ is the mass of the companion in solar masses.
For a system with $P_{\rm orb} > P_{\rm crit}$ (e.g.\ $P_{\rm orb} = 3$ h in 
Fig. 1), once a drop of the 
mass-transfer rate has started the radio ejection, a subsequent restoration 
of the original mass-transfer rate is unable to quench the ejection process
(see Burderi \etal 2001 for a detailed discussion of this effect), thus
prolonging the ejection phase. 

This ``radio-ejection'' phase envisaged by Burderi \etal (2001),
has been spectacularly demonstrated by the discovery of PSR~J1740-5340,
an eclipsing MSP, with a spin period of 3.65 ms and an orbital period 
of 32.5 h (D'Amico \etal 2001), very close to $P_{\rm crit}$ for the 
parameters of this system (Burderi, D'Antona, \& Burgay 2002).
The peculiarity of this system is that the companion is still overflowing 
its Roche lobe. This is demonstrated by the presence of
matter around the system that causes the long lasting and sometimes
irregular radio eclipses, and by the shape of the optical light curve,
which is well modeled assuming a Roche-lobe deformation of the mass
losing component (Ferraro \etal 2001).  An evolutionary scenario for
this system has been proposed by Burderi \etal (2002),
who provided convincing evidence that \psr is an example of a system
in the radio-ejection phase, by modeling the evolution of the
possible binary system progenitor.

In conclusion, while the evolution without a radio-ejection phase
implies that a large fraction of the transferred mass is accreted onto
the NS (because of the constraints imposed by the virial theorem), we
have demonstrated that the switch-on of a radio pulsar (associated to
a significant drop in mass transfer) could determine long episodes 
characterized by ejection
efficiencies close to 100\%, as the matter is ejected before it falls
into the deep gravitational potential well of the primary. This phenomenon
can prevent gravitational collapse and ultra-fast spinning NSs.

\end{document}